\documentstyle[sprocl]{article}

\def\be{\begin{equation}}
\def\ee{\end{equation}}
\def\bea{\begin{eqnarray}}
\def\eea{\end{eqnarray}}

\newcommand{\lsim}{\mathrel{\lower4pt\hbox{$\sim$}}
\hskip-12.5pt\raise1.6pt\hbox{$<$}\;}

\newcommand{\gsim}{\mathrel{\lower4pt\hbox{$\sim$}}
\hskip-12.5pt\raise1.6pt\hbox{$>$}\;}

\begin{document}

\rightline{BNL-HET-00/3}
\vskip 1.5cm

\title{Hadronic matrix elements from the lattice
\footnote{To appear in Proceedings of The Third International Conference on
B Physics and CP Violation, Taipei, December 3$-$7, 1999, H. -Y. Cheng and
W. -S. Hou, eds. (World Scientific, 2000).}} 

\author{Amarjit Soni}

\address{Theory Group, Brookhaven National Laboratory, Upton, NY\ \
11973, USA \\E-mail: soni@bnl.gov}

\maketitle

\abstracts{Lattice matrix elements are briefly reviewed. In the quenched
approximation $f_B$, $B_B$ and $B_K$ are now under good control.
Experimental hints for $f^{\rm expt}_{D_S}> f^{\rm QQCD}_{D_S}$ are
noted; precise determination of $f_{D_S}$ from experiment as well as
from the lattice is strongly advocated. Lattice calculations of the
form factor for $B\to\pi\ell\nu$ at relatively large value of $q^2$
have made good progress and should be useful in conjunction with
precise measurement of the differential spectra expected from
$B$-factories. Recent attempt at $K\to \pi\pi$ using staggered quarks
is briefly discussed; use of non-perturbative renormalization, improved
actions and operators with staggered quarks is emphasized. Due to the
good chiral behavior of domain wall quarks it would be useful
to study $K\to
\pi\pi$ with this discretization.}

\section{Introduction}

Hadronic matrix elements are of crucial importance for constraining the
parameters of the Standard Model (SM) in conjunction with experimental
information. Lattice approach has already attained considerable success
in handling $B$-parameters and decay constants and to a lesser degree
semi-leptonic form factors.~\cite{heavy} Weak decays into purely
hadronic final states, of which $K\to\pi\pi$ is the simplest example,
are extremely problematic.~\cite{kaon} After almost a decade the topic
is again getting some attention due in part to important developments
with regard to chiral symmetry on the lattice.~\cite{blum}

\section{Heavy-light decay constants: effects of
quenching.~\protect\cite{heavy}}

Lattice methods have made considerable progress in pinning down, fairly
accurately, $f_B$ and other heavy-light decay constants, in the
quenched approximation (see Table~\ref{tabone}). Unfortunately, the
effects of quenching could be substantial; current estimates place them
around $\sim (20\pm10)\%$, seriously limiting the phenomenological
applications. Since the past few years there is heightened activity to
accurately ascertain the effects of quenching. For technical reasons,
the dynamical simulations are with $N_f=2$ i.e.\ with only two
``light'' sea-quarks rather than the three ($u,d,s$) in real life.
Furthermore, the mass of the sea-quarks tends to be relatively heavy.
The indications from these dynamical simulations is that $f^{\rm
dynamical}_B > f^{\rm quenched}_B$.

\begin{table}[tbh]
\caption{Heavy-light decay constants and their
ratios.~\protect\cite{heavy} \label{tabone}}
\begin{center}
\begin{tabular}{|l|c|c|}
\hline
Quantity & Quenched ($N_f=0$) & Partially Unquenched ($N_f=2$) \\
\hline
$f_B$/MeV & $170\pm20$ & $200\pm30$ \\
$f_{B_S}$/MeV & $190\pm20$ & $220\pm30$ \\
$f_D$/MeV & $205\pm20$ & $225\pm30$ \\
$f_{D_S}$/MeV & $225\pm20$ & $245\pm30$ \\
$f_{B_S}/f_B$ & $1.14\pm.06$ & $1.14\pm.06$ \\
$f_{D_S}/f_D$ & $1.10\pm.06$ & $1.10\pm.06$ \\
\hline
\end{tabular}
\end{center}
\end{table}

\section{Hints from Experiment: $f^{\rm expt}_{D_S}>f^{\rm QQCD}_{D_S}$
?}

Table~\ref{tabtwo} exhibits a compilation of the results for $f_{D_S}$
from several experiments. Curiously, the central value of all but one
experiment is above the value for $f_{D_S}$ from quenched QCD
simulations: $f^{\rm QQCD}_{D_S} = 225\pm20$ MeV\null. Since the errors
in the existing experimental numbers are rather large we cannot draw
strong conclusions; it seems plausible, nevertheless, that these
experiments are also indicating that quenched QCD tends to
underestimate the heavy-light pseudoscalar decay constants.

\begin{table}[tbh]
\caption{Experimental determinations of $f_{D_S}$.~\protect\cite{aoki}
\label{tabtwo}}
\begin{center}
\begin{tabular}{|l|c|c|}
\hline
Experiment & Mode & $f_{D_S}$/MeV \\
\hline
S. Aoki {\it et al}. (WA75) & $\mu^+\nu_\mu$ &
\quad$238\pm47\pm21\pm48$~~\quad \\
J.Z. Bai {\it et al}. (BES) & {\tt "} & $430^{+150}_{-130}\pm40$ \\
K. Kodema {\it et al}. (E653) & {\tt "} & $190\pm34\pm20\pm26$ \\
M. Chada {\it et al}. (CLEO) & {\tt "} & $280\pm19\pm28\pm34$ \\
M. Acciari {\it et al}. (L3) & $\tau^+\nu_\tau$ & $309\pm58\pm33\pm38$
\\
F. Parodi {\it et al}. (DELPHI) & {\tt "} & $330\pm95$ \\
\phantom{F. Pareodi {\it et al}. }(ALEPH) & \quad inclusive~~\quad &
$284\pm62$ \\
\hline
\end{tabular}
\end{center}
\end{table}

\section{Precise studies of $f_{D_S}$}

For experiment as well as for dynamical lattice simulations, a precise
determination of $f_{B_D}(f_{B_S})$ is significantly more problematic
than $f_{D_S}$; in fact experimentally direct determination of $f_B$ is not
of immediate reach. Therefore, it would be very useful if the
experimental as well as the lattice determinations of $f_{D_S}$ could be
improved to 10--15\% accuracy. A comparison between the two would then
serve as a very useful guide for correcting $f_{B_d}(f_{B_S})$ from the
lattice. In this regard the ratios $f_{B_d}/f_{D_S}$ and
$f_{B_S}/f_{D_S}$ from the lattice would clearly be useful.

\section{Heavy-light $B$ parameters}

Lattice has had success with heavy-light $B$-parameters for a long time
although calculations in the static approximation are still somewhat
problematic.~\cite{heavy} Table~\ref{tabthree} presents a brief summary.
Recall that a
precise value of the ratio $f_{B_S} (B_{B_S})^{1/2}/f_{B_d}
(B_{B_d})^{1/2}$ is needed in accurately deducing $V_{td}/V_{ts}$
once $B_S$-$\bar B_S$ oscillations get experimentally measured.

\begin{table}[tbh]
\caption{Summary of heavy-light $B$-parameters.~\protect\cite{heavy}
\label{tabthree}}
\begin{center}
\begin{tabular}{|l|c|c|}
\hline
 & Quenched &  ``Unquenched''  \\
\hline
$B_{B_d}(m_b)$ & .86(4)(8) & .86(4)(8) \\
$B_{B_S}/B_{B_d}$ & 1.00(1)(2) & 1.00(1)(2) \\
$f_{B_d}(\hat B^{nlo}_{B_d})^{1/2}$ & $195\pm25$ MeV & $230\pm35$ \\
$\frac{f_{B_S}(B^{nlo}_{B_S})^{1/2}}{f_{B_d} (B^{nlo}_{B_d})^{1/2}}$ &
$1.14\pm.06$ & $1.14\pm.07$ \\
\hline
\end{tabular}
\end{center}
\end{table}

In view of the importance of the aforesaid ratio it may be useful to
determine it also more directly from the SU(3) breaking
ratio:~\cite{bern}

\[
\frac{M_{B_S}(\mu)}{M_{B_d}(\mu)} = \frac{\langle \bar B_{B_S} | (\bar
b\gamma_\rho (1-\gamma_5)s)^2| B_{B_S} \rangle}{\langle \bar B_{B_d}|
(\bar b\gamma_\rho(1-\gamma_5)d)^2 | B_{B_d}\rangle}
\]

\noindent Many of the systematic errors should cancel in this ratio.
However, in the attempts that have been so far made with this direct
method,~\cite{bern,lell} the errors are not yet small enough to make
this method competitive with the traditional $f^2_BB$ method.

\section{Semi-leptonic form factors, $B\to\pi(\rho)\ell\nu$.}

The large $b$-quark mass presents a difficult computational problem.
The cleanest lattice simulations are for `rest to rest' i.e.\ both the
initial and final meson  at rest. Then the large $B$ mass forces
$q^2$ ($q={}$lepton 4-momentum${}=p_B-p_\pi$) to be rather large. For
some phenomenological applications the value of the form factor(s) are
needed near $q^2\sim0$. This entails large extrapolations, introducing
additional errors and model dependence.  However,  one important
phenomenological application, namely deducing the mixing angle $V_{ub}$
from experimental data, only requires precise knowledge of the form
factor at one value of $q^2$. This should work so long as the experiment has
enough data
so that even the differential rate around that region of $q^2$ is
accurately determined, which is anticipated to be feasible at the
$B$-factories. In this regard two recent
approaches are noteworthy. Both of these efforts avoid the use of large
extrapolations in $q^2$ and focus instead on accurate predictions in a
limited region of $q^2$.

UKQCD focussed on near the end-point or the zero-recoil region where
the lattice data tends to be cleanest. Furthermore, heavy quark
symmetry also provides useful scaling relations in this
region.~\cite{isgur} Their result for the form factor $f^+$ as a
function of $q^2$ is given below.

\begin{table}[tbh]
\caption{$f^+(q^2)$ from UKQCD.~\protect\cite{bowl}
\label{tabfour}}
\begin{center}
\begin{tabular}{|l|c|c|c|c|c|}
\hline
$q^2$(GeV$^2$) & 16.7 & 18.1 & 19.5 & 20.9 & 22.3 \\
$f^+(q^2)$ & $0.9^{+1+2}_{-2-1}$ & $1.1^{+2+2}_{-2-1}$ &
$1.4^{+2+3}_{-2-1}$ & $1.8^{+2+4}_{-2-1}$ & $2.3^{+3+6}_{-3-1}$ \\
\hline
\end{tabular}
\end{center}
\end{table}

The FNAL group~\cite{ryan} is also focussing on an approach towards the
semi-leptonic form factors for $B\to\pi\ell\nu$, $D\to \pi(K)\ell\nu$
suitable for accurate determinations of mixing-angles in conjunction
with high statistics data samples anticipated from experiments. The key
idea here again is to concentrate directly on the differential decay
spectrum in an interval with $0.4\lsim \vec p_\pi/{\rm GeV}\lsim0.8$
thus avoiding the need for large extrapolation in $q^2$. The partial width
over this interval can be computed on the lattice. From the
experimental point of view this should have the advantage of using a
range of $q^2$ wherein the decay rate is not small unlike near the
end-point.

\section{$B_K$}

The kaon $B$-parameter, $B_K$, has been studied most extensively on the
lattice.~\cite{kaon} Two important limitations that lattice simulations
still need
to adequately address are SU(3) breaking (the `kaon' on the lattice is
a pseudoscalar made of degenerate quarks with $m_{\rm pseudoscalar}\sim
m_K$) and the quenched approximation. Both of these effects are expected
to be rather small $\sim5$--10\% and we get $\hat B_K=.85\pm.13$. This
number is
based on results of various groups ~\cite{kaon}, amongst which the one from
JLQCD is
the most precise.~\cite{aokitwo} The error on the lattice number
contains a guess-estimate of the SU(3) breaking and quenching errors.
We note, in passing, that there are some preliminary indications that
unquenching will increase~\cite{kilcup} or decrease~\cite{soni} $B_K$
by just a few percent. For now,
we have  not changed the central value of $B_K$ due to this
effect; only the systematic errors are increased to reflect this
possibility.

\section{$K\to2\pi$ Decays and $\epsilon^\prime/\epsilon$.}

It was realized long ago that chiral perturbation theory (ChPT) can be
used to simplify the problem so that $\langle \pi\pi|Q|K\rangle$ can be
obtained by computing on the lattice simpler entities:
$\langle\pi|Q|K\rangle$ and $\langle{\rm vac}|Q|K\rangle$, where $Q$ is
a 4-quark operator.~\cite{berntwo} The coefficients in this relation
can be calculated using lowest order ChPT\null. Traditionally this
strategy has been available for staggered fermions as they possess a
remnant chiral symmetry.~\cite{kiltwo} Since Wilson fermions explicitly
break chiral symmetry this approach cannot be used with this
discretization.~\cite{bochi} The new development in this regard is that
domain wall
quarks (DWQ)~\cite{kaplan} are found to be quite practical for
simulating QCD and possess excellent chiral behavior.~\cite{blumtwo}
Therefore, the $K\to\pi$ (and $K\to{}$vac) method for dealing with
$K\to2\pi$ is amenable to this discretization as well.~\cite{blum}

Final state interactions (FSI) are of course a serious limitation of
the $K\to\pi$ method.
However, it should still be very instructive to quantify how well this
concrete approximation works. Actually direct $K\to2\pi$ decays may
also be amenable to the lattice due to an elegant application of the
CPS symmetry.~\cite{bernthree} The restrictions of the
Maiani-Testa theorem~\cite{mai} are bypassed here by working near
the threshold. In any case the direct $K\to2\pi$ methods are
computationally extremely intensive and for now there is not much to
report based on these methods.

The credit for an extensive study of the $K\to2\pi$ in the $K\to\pi$
approach with staggered fermions goes to Pekurovsky and Kilcup
(PK).~\cite{peku,pekutwo} The work presents a first study of lattice
spacing dependence, finite size effects as well as quenching effects.
Unfortunately PK used lattice weak coupling perturbation
theory~\cite{sharpe} (LWCPT) to renormalize the operators and this seems to
completely fail for staggered quarks in the case of LR operators such
as $Q_6$ which are
crucial for $\epsilon^\prime/\epsilon$.

At $\beta=6.0$, PK observe a significant enhancement of the $\Delta
I=1/2$ channel over the 3/2; however at $\beta=6.2$, i.e.\ closer to
the continuum limit (compared to $\beta=6.0$), they find that the
central value of the enhancement weakens appreciably. While the large
errors at $\beta=6.2$ do not allow a strong conclusion, more work is
needed to unambiguously show that the $\Delta I=1/2$ enhancement
survives in the continuum limit.

PK's calculation of $\epsilon^\prime/\epsilon$ is seriously hampered as
the one-loop LWCPT that they use for renormalization completely fails
for $Q_6$. The perturbation theory corrections are several hundreds of
percents showing extreme sensitivity to the renormalization scale as
well as the quark mass (see Table~\ref{tabfive}). As PK themselves clearly
emphasize their
calculation of $\epsilon^\prime/\epsilon$ is extremely fragile and not
at all reliable.

\begin{table}[tbh]
\caption{$\langle Q_6\rangle$ in arbitrary units with one-loop
perturbative matching using two values of $q^\ast$; for comparison, the
renormalized results (``bare'') are also given (Table 2 from Ref. 20).
\label{tabfive}}
\begin{center}
\footnotesize
\begin{tabular}{|l|c|c|c|c|c|}
\hline
Quark Mass & 0.01 & 0.02 & 0.03 & 0.04 & 0.05 \\
\hline
$q^\ast=1/a$ & $0.1\pm1.2$ & $-0.9\pm0.4$ & $-1.2\pm0.2$ & $-1.6\pm0.3$
& $-1.1\pm0.2$ \\
$q^\ast=\pi/a$ & $-13.1\pm1.8$ & $-9.0\pm0.5$ & $-7.1\pm0.3$ &
$-6.3\pm0.5$ & $-4.6\pm0.5$ \\
Bare & $-55.6\pm5.0$ & $-35.4\pm1.5$ & $-27.0\pm0.9$ & $-22.3\pm1.4$ &
$-16.4\pm1.5$ \\
\hline
\end{tabular}
\end{center}
\end{table}

It seems quite plausible that for the renormalization of the $\Delta
S=1$, 4-quark operators, as has also been known to some extent to be  the
case for quark bilinears, the breaking of flavor symmetry by the
staggered approach is responsible for the failure of LWCPT\null. In any
case, use of non-perturbative renormalization (NPR)~\cite{marti}
improved actions and/or operators with the staggered approach are
highly desirable in the context of $K\to\pi\pi$ decays.

Domain wall quarks are extremely attractive as at the expense of
introducing a fictitious $5^{th}$ dimension they preserve the full
SU($N)\times{}$SU($N$) chiral symmetries of the continuum theory in the
limit of an infinite $5^{th}$ dimension.~\cite{kaplan} Quenched QCD
numerical simulations showed that in practice for $\beta\gsim6.0$,
10--20 sites in the $5^{th}$ dimension may be sufficient to render very
good chiral behavior.~\cite{blumtwo} Early numerical studies also seem
to indicate that the discretization errors are effectively $0(a^2)$; if
substantiated this improved scaling behavior may off-set the extra
cost of the $5^{th}$ dimension.~\cite{blumtwo,kiku}

Calculation of $K\to2\pi$ and $\epsilon^\prime/\epsilon$ in this method
has been in progress for quite sometime.~\cite{blum} With DWQ,
considerable progress has been made in non-perturbative
renormalization of quark bilinears, $\Delta S=2$ and $\Delta S=1$
Hamiltonians and so far the method seems promising.~\cite{dawson}

\section*{Acknowledgments}
I thank the organizers for a very interesting workshop. This work was
supported in part by the U.S. DOE contract DE-AC02-98CH10886.
\bigskip

\def \ajp#1#2#3{Am. J. Phys. {\bf#1}, #2 (#3)}
\def \apny#1#2#3{Ann. Phys. (N.Y.) {\bf#1}, #2 (#3)}
\def \app#1#2#3{Acta Phys. Polonica {\bf#1}, #2 (#3)}
\def \arnps#1#2#3{Ann. Rev. Nucl. Part. Sci. {\bf#1}, #2 (#3)}
\def \art{and references therein}
\def \cmts#1#2#3{Comments on Nucl. Part. Phys. {\bf#1}, #2 (#3)}
\def \cn{Collaboration}
\def \cp89{{\it CP Violation,} edited by C. Jarlskog (World Scientific,
Singapore, 1989)}
\def \epjc#1#2#3{Euro.~Phys.~J.~C {\bf #1}, #2 (#3)}
\def \epl#1#2#3{Europhys.~Lett.~{\bf #1}, #2 (#3)}
\def \ib{{\it ibid.}~}
\def \ibj#1#2#3{~{\bf#1}, #2 (#3)}
\def \ijmpa#1#2#3{Int. J. Mod. Phys. A {\bf#1}, #2 (#3)}
\def \jpb#1#2#3{J.~Phys.~B~{\bf#1}, #2 (#3)}
\def \jhep#1#2#3{JHEP {\bf#1}, #2 (#3)}
\def \mpla#1#2#3{Mod. Phys. Lett. A {\bf#1}, #2 (#3)}
\def \nc#1#2#3{Nuovo Cim. {\bf#1}, #2 (#3)}
\def \npb#1#2#3{Nucl. Phys.B {\bf#1}, #2 (#3)}
\def \npps#1#2#3{Nucl. Phys. Proc. Suppl. {\bf#1}, #2 (#3)}
\def \pisma#1#2#3#4{Pis'ma Zh. Eksp. Teor. Fiz. {\bf#1}, #2 (#3) [JETP Lett.
{\bf#1}, #4 (#3)]}
\def \pl#1#2#3{Phys. Lett. {\bf#1}, #2 (#3)}
\def \pla#1#2#3{Phys. Lett. A {\bf#1}, #2 (#3)}
\def \plb#1#2#3{Phys. Lett. B {\bf#1}, #2 (#3)}
\def \pr#1#2#3{Phys. Rev. {\bf#1}, #2 (#3)}
\def \prc#1#2#3{Phys. Rev. C {\bf#1}, #2 (#3)}
\def \prd#1#2#3{Phys. Rev. D {\bf#1}, #2 (#3)}
\def \prl#1#2#3{Phys. Rev. Lett. {\bf#1}, #2 (#3)}
\def \prp#1#2#3{Phys. Rep. {\bf#1}, #2 (#3)}
\def \ptp#1#2#3{Prog. Theor. Phys. {\bf#1}, #2 (#3)}
\def \ptwaw{Plenary talk, XXVIII International Conference on High Energy
Physics, Warsaw, July 25--31, 1996}
\def \rmp#1#2#3{Rev. Mod. Phys. {\bf#1}, #2 (#3)}
\def \rp#1{~~~~~\ldots\ldots{\rm rp~}{#1}~~~~~}
\def \stone{{\it $B$ Decays} (Revised 2nd Edition), edited by S. Stone
(World Scientific, Singapore, 1994)}
\def \yaf#1#2#3#4{Yad. Fiz. {\bf#1}, #2 (#3) [Sov. J. Nucl. Phys. {\bf #1},
#4 (#3)]}
\def \zhetf#1#2#3#4#5#6{Zh. Eksp. Teor. Fiz. {\bf #1}, #2 (#3) [Sov. Phys. -
JETP {\bf #4}, #5 (#6)]}
\def \zpc#1#2#3{Zeit. Phys. C {\bf#1}, #2 (#3)}
\def \zpd#1#2#3{Zeit. Phys. D {\bf#1}, #2 (#3)}

\end{document}